\documentclass[12pt]{article}
\usepackage{amsfonts}
\usepackage{amsmath}
\usepackage{amssymb}
\usepackage{amscd}
\usepackage[dvips]{graphicx}

\begin{document}

\begin{center}
%\section*
{\Large NONCOMMUTATIVE SPACE-TIME FROM QUANTIZED TWISTORS }
\end{center}

\bigskip

\begin{center}
{{\large ${\mathrm{Jerzy\;Lukierski}}$, ${\mathrm{Mariusz\;Woronowicz}}$ }}

\bigskip

{%\large
$\mathrm{~Institute\;of\;Theoretical\;Physics}$}

{%\large
$\mathrm{\ University\; of\; Wroclaw\; pl.\; Maxa\; Borna\; 9,\; 50-206\;
Wroclaw,\; Poland}$}

{%\large
$\mathrm{\ e-mail:\;lukier@ift.uni.wroc.pl;woronow@ift.uni.wroc.pl}$}

\bigskip
\end{center}
\begin{abstract}
We consider the relativistic phase space coordinates $(x_{\mu },p_{\mu })$
as composite, described by functions of the primary pair of twistor
coordinates. It appears that if twistor coordinates are canonicaly quantized
the composite space-time coordinates are becoming noncommutative. We obtain
deformed Heisenberg algebra which in order to be closed should be enlarged
by the Pauli-Lubanski four-vector components. We further comment on
star-product quantization of derived algebraic structures which permit to
introduce spin-extended deformed Heisenberg algebra.
\end{abstract}

\section{Introduction}

\bigskip Space-time description of relativistic point particles does not
provide a natural geometrization of spin degrees of freedom. It is well
acknowledged however that spin degrees of freedom play essential role in the
description of space-time as dynamical system, what is illustrated e.g. by
spin foam approaches to quantum gravity (see e.g. \cite{1},\cite{2})or the
use of spin networks in loop quantum gravity (see e.g. \cite{3},\cite{4}).
Well known geometrization of the spin degrees of freedom is provided by
superspace extensions of space-time (see e.g. \cite{5},\cite{6}), with
finite-dimensional Grassmann algebra attached to each space-time point. In
this paper we shall introduce geometric spin degrees in different way by
considering as primary the twistor geometry (see e.g. \cite{7},\cite{8})
with basic spinorial coordinates, and consider space-time coordinates as
their composites.

The twistors in $D=4$ are the fundamental conformal $SU(2,2)$ spinors and
introduce primary conformal geometry, with single twistors well suited to
the description of massless elementary objects \cite{7}-\cite{9} In fact
single twistor space has exactly the structure of phase space for massless
particle, with all possible choices of helicity \cite{10}. Massive particles
with arbitrary spin can be described if we introduce two-twistor space,
which contains the phase space for massive particles with spin \cite{9},\cite%
{11}-\cite{14}. The pair of twistors is needed as well if we wish to
introduce the space-time coordinates given as composites of twistor
components \cite{7}. In this report we shall consider canonically quantized
pairs of twistors which provide particular choice of noncommutative
composite space-time coordinates and leads to the generalization of the
standard QM phase space structure.

In order to introduce the coordinates describing point in complex Minkowski
space-time one should employ two nonparalell twistors $t_{A,i}$ where $%
A=1,2,3,4$ are the $SU(2,2)\simeq O(4,2)$ indices, and $i=1,2$ is the
internal $U(2)$ index. In two-twistor complex space $T^{(2)}=T\otimes T\in
t_{A,i}$ ($t_{A,1}\in T\otimes 1,$ $t_{A,2}\in 1\otimes T$) one can
introduce the following canonical twistorial Poisson brackets (PB) \cite{7}%
\begin{equation}
\{\bar{t}^{A,i},t_{\,B,j}\}=\delta _{B}^{A}\delta _{j}^{i}\qquad \{\bar{t}%
_{i}^{A,i},\bar{t}_{\,}^{B,j}\}=\{t_{A,i},t_{\,B,j}\}=0  \label{1}
\end{equation}%
where $\bar{t}^{A,i}=g^{AB}$ $\bar{t}_{B}^{\quad i}$ and $g^{AB}$ describes
the Hermitean $SU(2,2)$ metric. The standard choice of twistor coordinates,
described by the pairs of $2$-component Weyl spinors $(t_{A,i}=(\pi _{\alpha
,i},\omega _{\quad i}^{\dot{\alpha}}))$, corresponds to the metric $%
g^{AB}=\left( 
\begin{array}{cc}
0 & 1 \\ 
1 & 0%
\end{array}%
\right) $.

One can introduce the bilinear Hermitean products of twistors $t_{A,i},$ $%
\bar{t}_{A}^{\quad i}$ which after the use of (\ref{1}) describe the
twistorial realization of conformal algebra $O(4,2)$ \cite{7}. In particular
for the Poincare algebra generators $P_{\mu }=P_{\dot{\alpha}\beta },$ $%
M_{\mu \nu }=(M_{\alpha \beta },M_{\dot{\alpha}\dot{\beta}})$ \ we get the
formulae%
\begin{align}
\ P_{\alpha \dot{\beta}}& =\pi _{\alpha }^{i}\overline{\pi }_{\dot{\beta}%
,i}\qquad  \label{2} \\
\qquad M_{\alpha \beta }& =\omega _{(\alpha }^{i}\pi _{\beta )}^{i}\qquad M_{%
\dot{\alpha}\dot{\beta}}=\overline{\omega }_{(\dot{\alpha}}^{i}\overline{%
\omega }_{\dot{\beta})}^{i}.  \label{3}
\end{align}%
The complex Minkowski space-time coordinates parametrize two-planes in
twistor space, which are determined by the pair of incidence equation for
two twistors $t_{A}^{i}=(\pi _{\alpha }^{i},\omega ^{\dot{\alpha}i})$%
\begin{equation}
\ \omega _{\;}^{i\dot{\beta}}=\pi _{\alpha }^{i}z^{\alpha \dot{\beta}}\qquad
z^{\overset{\cdot }{\alpha }\beta }=(\sigma _{\mu })^{\overset{\cdot }{%
\alpha }\beta }z^{\mu }  \label{4}
\end{equation}%
which provides the known composite formula for $z^{\alpha \dot{\beta}}$ \cite%
{7}-\cite{9}%
\begin{equation}
z^{\alpha \dot{\beta}}=\frac{i}{\pi ^{1\alpha }\pi _{\alpha }^{2}}(\pi
^{1\alpha }\omega ^{2\dot{\beta}}-\pi ^{2\alpha }\omega ^{1\dot{\beta}%
})=x^{\alpha \dot{\beta}}+iy^{\alpha \dot{\beta}}.  \label{5}
\end{equation}%
One chooses that real part $x^{\alpha \dot{\beta}}=\Re z^{\alpha \dot{\beta}%
} $ describes the composite real \textit{physical} Minkowski space.
Twistorial PB (\ref{1}) induce further on the composite space-time
coordinates the noncommutative structure.

We mention that the relations (\ref{4})-(\ref{5}) has been already used as
defining the coordinates of quantum free fields (see e.g. \cite{e1}). Our
aim here is to incorporate the nonvanishing PB of composite Minkowski
space-time coordinates $x_{\mu }=\frac{1}{2}(\sigma _{\mu })_{\alpha \dot{%
\beta}}x^{\alpha \dot{\beta}}$ into enlarged deformed relativistic
Heisenberg algebra. It appears that in order to get the closure of PB
algebra it is necessary to add to space-time coordinates $x_{\mu }=\Re
z_{\mu }$\ (see (\ref{5})) and fourmomenta $p_{\mu }$ (see (\ref{2})) the
Pauli-Lubanski fourvector components $(\mu =0,1,2,3)$%
\begin{equation}
\ W_{\mu }=\frac{1}{2}\epsilon _{\mu \nu \rho \tau }P^{\nu }M^{\rho \tau }.
\label{6}
\end{equation}%
After insertion in (\ref{6}) of composite formulae (\ref{2})-(\ref{3}) we
obtain Pauli-Lubansky fourvector as expressed fourlinearly in twistor
coordinates $t_{A,i},$ $\bar{t}_{A}^{\quad i}$. In such a way we obtain
spin-extended deformed Heisenberg (SEDH) algebra with the basis described by
the generators ($x_{\mu },P_{\mu },W_{\mu }$).

\section{PB structure of spin-extended deformed Heisenberg algebra}

Let us introduce the Hermitean $2\times2$ matrix described by the $U(2,2)-$%
invariant scalar products $(r=1,2,3)$%
\begin{equation}
\ K_{j}^{i}=\bar{t}^{A,i}t_{A,j}=(K_{i}^{j})^{\dagger}=(%
\tau_{r})_{j}^{i}k_{r}+\delta_{j}^{i}k_{0}  \label{7}
\end{equation}
where $\tau_{r}$ are Pauli matrices and $\bar{k}_{r}=k_{r}$, $\bar{k}%
_{0}=-k_{0}$. The scalar products $K_{j}^{i}$ introduce the internal $U(2)$
PB algebra induced by the canonical twistorial PB (\ref{1}):%
\begin{equation}
\ \{k_{r},k_{s}\}=\epsilon_{rst}k_{\tau}\qquad\{k_{0},k_{r}\}=0.  \label{8}
\end{equation}

In order to describe the fourlinear twistor formula \ for $W_{\mu }$in
compact way one can introduce four composite vierbeins by means of the
formula (see also \cite{15})%
\begin{equation}
\ e_{\mu }^{(\rho )}=\frac{1}{2}(\sigma _{\mu })^{\alpha \dot{\beta}}%
\overline{\pi }_{\dot{\alpha}}^{i}(\tau ^{\rho })_{i}^{j}\pi _{\alpha j}
\label{9}
\end{equation}%
where $\tau ^{(\rho )}=(1_{2},\tau _{r})$. One obtains from the comparison
with (\ref{2}) that $e_{\mu }^{(0)}=p_{\mu }$ and one gets the ortogonality
relations%
\begin{equation}
\ e_{\mu }^{(\rho )}e^{\mu (\tau )}=|f|^{2}\eta ^{\rho \tau },\qquad \ f=\pi
_{\alpha }^{1}\pi ^{\alpha 2}\qquad |f|^{2}=p^{2}.  \label{10}
\end{equation}%
Due to (\ref{10}) the set of composite frame fields $\ e_{\mu }^{(\rho )}$\
depends on seven independent degrees of freedom which describe eight degrees 
$(\pi _{\alpha ,i},\overline{\pi }_{\dot{\alpha}}^{\quad i})$ \ factorized
by $U(1)$ phase $\pi _{\alpha ,i}\rightarrow e^{i\gamma }\pi _{\alpha ,i}$.
Further one can derive the formulae (see also \cite{6})%
\begin{equation}
\ W_{\mu }=k_{r}e_{\mu }^{(r)}\qquad k_{r}=-\frac{1}{p^{2}}e_{\mu
}^{(r)}W^{\mu }  \label{12}
\end{equation}%
or more explicitely $(W_{\alpha \dot{\beta}}=(\sigma ^{\mu })_{\alpha \dot{%
\beta}}W_{\mu })$%
\begin{equation}
\ W_{\alpha \dot{\beta}}=k_{3}(\pi _{\alpha }^{1}\overline{\pi }_{\dot{\beta}%
}^{1}-\pi _{\alpha }^{2}\overline{\pi }_{\dot{\beta}}^{2})+k_{+}\pi _{\alpha
}^{2}\overline{\pi }_{\dot{\beta}}^{1}+k_{-}\pi _{\alpha }^{1}\overline{\pi }%
_{\dot{\beta}}^{2}  \label{13}
\end{equation}%
where $k_{\pm }=k_{1}\pm ik_{2}$.

The relations (\ref{12}) provide the covariant formulae for the generators $%
k_{r}$ of internal $SU(2)$ algebra (see (\cite{7})) which describe the
Lorentz-invariant three spin projections. From (\ref{10})-(\ref{12}) foloows
that%
\begin{equation}
\ W_{\mu}W^{\mu}=p^{2}t^{2}\qquad t^{2}=k_{1}^{2}+k_{2}^{2}+k_{3}^{2}
\label{14}
\end{equation}
and after quantization of PB (\ref{8}) we obtain the well-known relativistic
spin square spectrum with $t^{2}$ replaced by quantum spin square $s(s+1)$ $%
(s=0,\frac{1}{2},1,\ldots)$.

We see from (\ref{12})-(\ref{13}) that the component $k_{0}$ does not enter
into the definition of composite Pauli-Lubansky fourvector, but one can show
that it contributes to the imaginary part $y_{\mu }=\Im z_{\mu }$ of
composite complex Minkowski space coordinates (\ref{5}). One can derive the
following general formula 
\begin{equation}
\ y_{\mu }=-\frac{1}{p^{2}}k^{\rho }e_{\mu }^{(\rho )}=-\frac{1}{p^{2}}%
(t^{0}p_{\mu }-W_{\mu }).  \label{15}
\end{equation}

The choice simplyfying the formulae (see (\ref{13})) for the spin fourvector 
$W_{\mu }$ and complex Minkowski space are obtained if we choose $%
t_{3}=t\neq 0$ and $t_{0}=t_{1}=t_{2}=0$. We obtain that%
\begin{equation}
\ W_{\alpha \dot{\beta}}=t(\pi _{\alpha }^{1}\overline{\pi }_{\dot{\beta}%
}^{1}-\pi _{\alpha }^{2}\overline{\pi }_{\dot{\beta}}^{2}),\qquad \ z_{\mu
}=\ x_{\mu }+\frac{i}{p^{2}}W_{\mu }.  \label{17}
\end{equation}

Using explicite formulae (\ref{2}), (\ref{5}), (\ref{8}) and (\ref{13}) one
can derive the following PB algebra (see also \cite{12}; we denote further
the twistor functions (\ref{2}) and (\ref{13}) by small letters)%
\begin{align}
\{x_{\mu },p_{\nu }\}& =\eta _{\mu \nu }\qquad \{p_{\mu },p_{\nu }\}=0
\label{19} \\
\{x_{\mu },x_{\nu }\}& =-\frac{1}{(p^{2})^{2}}\epsilon _{\mu \nu \rho \sigma
}w^{\rho }p^{\sigma }  \label{20} \\
\{w_{\mu },x_{\nu }\}& =-\frac{1}{p^{2}}w_{[\mu }p_{\nu ]}\qquad \{w_{\mu
},p_{\nu }\}=0  \label{21} \\
\{w_{\mu },w_{\nu }\}& =\epsilon _{\mu \nu \rho \sigma }w^{\rho }p^{\sigma },
\label{22}
\end{align}%
\bigskip which are consitent with the relation $p_{\mu }W^{\mu }=0.$

The nonpolynomial PB algebra (\ref{19})-(\ref{22}) is consistent with
dimensionalities $[x_{\mu }]=m^{-1},$ $[p_{\mu }]=[w_{\mu }]=m$. We mention
that the PB subalgebra with generators ($p_{\mu },w_{\mu }$) was e.g.
studied in \cite{16}\ (see Appendix I) as relativistic spin algebra. We add
that the nonpolynomial factor $\lambda ^{2}=p^{-2}$ can not be replaced by
constant inverse mass square because of the following nonvanishing PB%
\begin{equation}
\{x_{\mu },\frac{1}{p^{2}}\}=-2\frac{1}{(p^{2})^{2}}p_{\mu }\quad
\Rightarrow \quad \{x_{\mu },\lambda ^{2}\}=-2\lambda ^{4}p_{\mu }.
\label{23}
\end{equation}

\section{Quantization of spin-extended deformed Heisenberg algebra}

\bigskip

The general Poisson brackets can be quantized if we use Kontsevich
quantization method \cite{17}-\cite{18} which solved the problem of
existence of associative $\star $-product representing multiplication on
quantized Poisson manifold. By this method naive quantization of phase space
functions via replacement $\{\cdot ,\cdot \}\rightarrow \frac{i}{\hbar }%
[\cdot ,\cdot ]$ has been modified in a way which leads to the validity of
Jacobi identities. The star product representation of the algebra describing
quantized Poisson brackets is obtained after performing the Weyl map of
elements of SEDH algebra%
\begin{equation}
\ f(\widehat{Y}_{a})\overset{W}{\longrightarrow }f(Y_{a})\qquad
Y_{a}=(x_{\mu },p_{\mu },w_{\mu },\Lambda )  \label{25}
\end{equation}%
and introducing the following homomorphic map of the products%
\begin{equation}
\ f(\widehat{Y}_{a})\cdot g(\widehat{Y}_{b})\overset{W}{\longrightarrow }%
f(Y_{a})\star g(Y_{b})  \label{26}
\end{equation}%
where $\Pi _{j}$ are bidifferential operators maximally of $2j$ order and%
\begin{align}
\Pi _{0}(f,g)& =f\cdot g\qquad \qquad  \label{27} \\
\Pi _{1}(f,g)-\Pi _{1}(g,f)& =\{f,g\}\quad  \label{28} \\
f(Y_{a})\star (g(Y_{a})\star h(Y_{a}))+cycl& =0.  \label{29}
\end{align}%
If the algebraic manifold with coordinates $Y_{a}$ is not a flat one in the
bidifferential operators $\Pi _{j}$ one should introduce suitably
covariantized derivatives \cite{20}.

The SEDH PB algebra (\ref{19})-(\ref{22}) $\{Y_{a},Y_{b}\}=\omega _{ab}(Y)$
can be introduced as dual to the $2$-form $\Omega _{2}$ 
\begin{equation}
\{Y_{a},Y_{b}\}=\omega _{ab}(Y)\overset{dual}{\longleftrightarrow }\Omega
_{2}=\omega ^{ab}(Y)dY_{a}\wedge dY_{b}  \label{30}
\end{equation}%
where $\omega _{ab}\omega ^{bc}=\delta _{a}^{c}$.

\bigskip

One can show that $\Omega _{2}$ in (\ref{30}) has the form (see (\cite{12}))%
\begin{equation}
\Omega _{2}=dp^{\mu }\wedge dx_{\mu }+\Omega _{2}^{Sour},  \label{32}
\end{equation}%
where%
\begin{equation}
\Omega _{2}^{Sour}=\frac{1}{2(p^{2})^{1/2}}\epsilon _{\mu \nu \rho \sigma
}w^{\rho }p^{\sigma }(\frac{1}{p^{2}}dp^{\mu }\wedge dp^{\nu }-\frac{1}{t^{2}%
}dw^{\mu }\wedge dw^{\nu })  \label{31}
\end{equation}

The two-form (\ref{32}) can be obtained from the canonical Liouville
one-form $\theta _{1}$ on $T\times T$ 
\begin{equation}
\theta _{1}=i(\overline{t}_{\text{ },i}^{A}\wedge dt_{A,i})=\frac{i}{2}%
(\omega _{\text{ },i}^{\alpha }d\pi _{\alpha ,i}+\overline{\pi }_{\dot{\alpha%
},i}d\overline{\omega }_{\text{ },i}^{\dot{\alpha}}-c.c.)  \label{ggg}
\end{equation}%
after introducing the coordinates $Y_{a}$ as functions (see (\ref{2}), (\ref%
{5}) and (\ref{8})) of the pair of twistor coordinates $t_{i}$ $\in T\times
T $. It should be added that the one-form (\ref{ggg}) pulled back on
one-dimensional trajectories ($\int \theta =\int dtL$) \ in generalized
phase space $Y_{a}=$($x_{\mu },p_{\mu },w_{\mu }$) defines the action of
massive particle with spin characterized by the fourvector $w_{\mu }$ (see 
\cite{14}) .

\section{Outlook}

\bigskip The basic PB structure (\ref{19})-(\ref{22}) \ of our extended
deformed phase space requires for its application consistent quantization.
In standard QM there is well-known Wigner formulation (see e.g. \cite{22})
realizing Weyl correspondence between quantum-mechanical operators and
phase-space classical functions. In the geometric scheme presented in this
paper the basic PB are more complicated, but fortunately one can quantize
them via Kontsevich star-product \cite{17}-\cite{19}. An important property
of the $\star $-quantization formula given by (\ref{26}) is its dependence
only on the Poisson structure function $\omega _{ab}(Y)$ (see (\ref{30}))
and its derivatives to all orders. In such a way one gets the modification
of naive correspondence rule between classical PB and quantum mechanical
phase space commutators in order to achieve the validity of Jacobi identity
(from naive quantization prescription one obtains nonvanishing Jacobiator
with leading $\hbar ^{2}$ term). We obtain the quantization rules for SEDH
algebra by calculating the perturbative formula for $\star $-product, which
takes for Poisson structure (\ref{30}) the following perturbative form in
the space of functions $f(Y_{a}),g(Y_{a})$ depending on spin-deformed
extended phase-space coordinates $Y_{a}$ (compare with (\ref{26})) \cite{17}%
\begin{align}
f\star g& =fg+\hbar \omega _{ab}f,_{a}g,_{b}+\frac{\hbar ^{2}}{2}\omega
_{ab}\omega _{cd}f,_{ac}g,_{bd}  \label{gfd} \\
& +\frac{\hbar ^{2}}{3}\omega _{ab}\omega
_{cd,b}(f,_{ac}g,_{d}-f,_{c}g,_{ad})+O(\hbar ^{3})  \notag
\end{align}%
iwhere $f\equiv f(Y_{a})$,$f,_{a}\equiv \frac{\partial }{\partial Y_{a}}%
f(Y_{a})$ etc. One can check that using (\ref{gfd}) the Jacobi relation (\ref%
{29}) is satisfied up to the $\hbar ^{3}$ terms.

Concluding, using Kontsevich $\hbar $-expansion of general $\star $-product
formula, the perturbative quantization of our PB (\ref{19})-(\ref{22}) can
be achieved. In such a way one gets the spin-extended deformed QM and one
can further test possible applications. In particular we conjecture that the
presence of additional spin coordinate $w_{\mu }$ can help to provide new
ways of describing the relativistic (stringy?) spin dynamics with infinite
set of spin values.

\bigskip

\textbf{Acknowledgements }

The paper has been supported financially by Polish National Science Centre
(NCN) No 2011/01/B/ST2/03354.

\end{document}